\documentclass[aps,prl,twocolumn,showpacs,superscriptaddress]{revtex4}
\usepackage{CJK}
\usepackage{bm}
\usepackage{graphicx}
\usepackage{color}
\usepackage{amsmath,amssymb,amsfonts,amsthm}

\def \TT {\textsf{T}}
\def \CC {\textsf{C}}
\def \PP {\textsf{P}}
\def \RR {\textsf{R}}

\def \AA {\textsf{A}}

\def \DD {\textsf{D}}
\def \EE {\textsf{E}}

\def \H {\mathcal{H}}
\def \K {\hat{\mathcal{K}}}

\def \T {\hat{T}}
\def \C {\hat{C}}
\def \P {\hat{P}}
\def \A {\hat{A}}
\def \B {\hat{B}}
\def \D {\hat{D}}
\def \E {\hat{E}}

\begin{document}

\title{Unified theory of $\PP\TT$ and $\CC\PP$ invariant topological metals and nodal superconductors}

\author{Y. X. Zhao}
\email[]{y.zhao@fkf.mpg.de}
\affiliation{Max-Planck-Institute for Solid State Research, D-70569 Stuttgart, Germany}
\affiliation{Department of Physics and Center of Theoretical and Computational Physics, The University of Hong Kong, Pokfulam Road, Hong Kong, China}

\author{Andreas P. Schnyder}
\email[]{a.schnyder@fkf.mpg.de}
\affiliation{Max-Planck-Institute for Solid State Research, D-70569 Stuttgart, Germany}

\author{Z. D. Wang}
\email[]{zwang@hku.hk}
\affiliation{Department of Physics and Center of Theoretical and Computational Physics, The University of Hong Kong, Pokfulam Road, Hong Kong, China}

\begin{abstract}
As $\PP\TT$ and $\CC\PP$ symmetries are fundamental in physics, we establish a unified topological theory of $\PP\TT$ and $\CC\PP$ invariant metals and 
nodal superconductors, based on the mathematically rigorous $KO$ theory. Representative models are constructed for all nontrivial  topological cases   in  dimensions $d=1,2$, and $3$, with their exotic physical meanings being elucidated in detail. 
Intriguingly, it is found that the topological charges of Fermi surfaces in the bulk determine an exotic direction-dependent distribution of  topological subgap modes on the boundaries.
Furthermore, by constructing an exact bulk-boundary correspondence, we show that the topological Fermi points of the   $\PP\TT$ and $\CC\PP$ invariant classes can appear as gapless modes on the boundary of topological insulators with a certain type of anisotropic crystalline symmetry.
\end{abstract}

\pacs{71.90.+q,03.65.Vf, 71.20.-b, 73.20.-r}

\maketitle

\textit{Introduction.--}
Since the discovery of topological insulators,  rapid progress has been made in our understanding of topological band theory~\cite{hasan:rmp,qi:rmp,chiu_review15}.
Recently, much attention has been paid to topological metals/semimetals, which are characterized by nodal band 
structures whose stability
is guaranteed by certain symmetries and a nontrivial wave function topology~\cite{Volovik:book,HoravaPRL05,ZhaoWangPRL13,matsuuraNJP13}. 
The Bloch wave functions of these gapless systems possess a nonzero topological invariant index (e.g., a Chern, winding, or $\mathbb{Z}_2$ number).
Thus far, several distinct types of topological semimetals have been explored, including semimetals with Dirac points~\cite{Cd3As2Chen2014,neupaneDiracHasan,XuHasanScience15}, Weyl points~\cite{wan_savrasov_PRB_11,Hosur_Weyl_develop,Ashvin_Weyl_review}, 
and Dirac line nodes~\cite{FangFu_PRB15,chiu_conf_series_15,chan_chiu_schnyder_arXiv_15,
kim_kane_rappe_PRL_15,yu_Ca3PdN_arXiv}. Experimentally, these topological phases are realized in many different systems.
For example,  Weyl point nodes have been reported to exist in TaAs~\cite{Xu_Weyl_2015_first,Weyl_discovery_TaAs} and NbAs~\cite{NbAs_Weyl_ARPES_nat_phys_15},
while Dirac line nodes occur in Ca$_3$P$_2$~\cite{xie_schoop_Ca3P2}, PbTaSe$_2$~\cite{cava_PbTaSe2_PRB_14,bian_hasan_arXiv_15}, ZrSiS~\cite{schoopZrSiS_arXiv15} and carbon allotropes~\cite{DiracLine,FanZhang2}.
Topological nodal phases have also been created artificially using photonic crystals~\cite{ling_joannopoulos_nat_photon_13,Lu_science_15}
and ultracold atoms in optical lattices~\cite{zhang_wang_arXiv_16,das_sarma_cold_atoms}.
Moreover, nodal topological band structures can arise in superconductors with unconventional pairing symmetries~\cite{FanZhang1,SchnyderJPCM15,Sato-CP}.

In exploring these nodal topological phases, 
several significant advances have been 
made recently to classify Fermi surfaces in terms of anti-unitary symmetries (e.g., time reversal and particle-hole symmetries)~\cite{ZhaoWangPRL13,matsuuraNJP13,ZhaoWangPRB14,zhao_wang_PRL_16},  as well as unitary symmetries (e.g., reflection and rotation symmetries)~\cite{ChiuSchnyder14,Sato_Crystalline_PRB14,PhysRevB.90.115207,yang_nagaosa_nat_commun_14,gaoZhangArxiv15}.
These pieces of work have broadened and deepened our knowledge of symmetry-protected topological materials.
However, a unified topological theory of nodal phases that possess the combined symmetry of unitary with anti-unitary operations, including a comprehensive classification,  is still badly awaited.
In particular, the combined symmetry of time-reversal \textsf{T} (or particle-hole  \textsf{C}) with inversion \textsf{P} is of fundamental importance.
Similar to particle physics, these combined symmetries  play fundamental roles in many condensed matter systems, such as centrosymmetric crystal structures, 
superfluid ${}^3$He~\cite{Volovik:book}, and possibly some heavy fermion superconductors~\cite{pfleiderer_RMP_09}.

In this Letter, we establish a unified theory  for the topological properties of $\PP\TT$ and $\CC\PP$ symmetry-protected nodal band structures, based on $KO$ theory~\cite{Atiyah-KR,Karoubi-book,Clifford-modules,Atiyahbook,KR-KO},
i.e., the K theory of real vector bundles. 
Using the homotopy groups of KO theory, we topologically classify $\PP\TT$ and $\CC\PP$ invariant Fermi surfaces (Table~\ref{Classification}).
Interestingly, in this classification the same K groups appear as in  the classification of strong topological insulators and superconductors (TIs and TSCs)~\cite{KR-KO,Kitaev2009,Schnyder2008,Ryu2010ten},
but in a reversed order. We construct concrete models for all topologically nontrivial Fermi surfaces in
dimensions $d=1$, $2$, and $3$,
and elaborate how the topological  charges of the Fermi surfaces in the bulk determine the distribution of topological subgap modes on the boundaries.
Furthermore, we show that the $\PP\TT$ and $\CC\PP$ invariant topological Fermi points can be reinterpreted as the boundary modes of TIs/TSCs with certain anisotropic crystalline symmetries 
[specified in Eqs.~(\ref{new_refl}) and (\ref{new_syms})], 
thereby realizing an exact bulk-boundary correspondence.

\textit{Topological Fermi surfaces with $\PP\TT$ symmetry.--}
Let us start by considering 
 systems with the combined symmetry $\PP\TT$. When a quantum system has both $\TT$ and $\PP$ symmetries, the Hamiltonian density $\mathcal{H}(k)$ in  momentum space satisfies 
\begin{equation}
\hat{T}\mathcal{H}(k)\hat{T}^{-1}=\mathcal{H}(-k) \;  \;  \textrm{and} \;  \; \hat{P}\mathcal{H}(k)\hat{P}^{-1}=\mathcal{H}(-k),
\end{equation} 
respectively, where $\hat{T}$ is an anti-unitary operator, $\hat{T}i\hat{T}^{-1}=-i$, while $\hat{P}$ is unitary, $\hat{P}i\hat{P}^{-1}=i$. In this work we only require the combined symmetry $\textsf{A}=\PP\TT$ for the topological stability, while perturbations breaking both $\TT$ and $\PP$ but preserving $\PP\TT$ are allowed. $\AA=\PP\TT$ is anti-unitary, $ \hat{A}i\hat{A}^{-1}=-i$, acting on $\mathcal{H}(k)$ as
\begin{equation}
\hat{A}\mathcal{H}(k)\hat{A}^{-1}=\mathcal{H}(k). 
\end{equation}
It is important to observe that the symmetry $\PP\TT$ operates trivially in  momentum space, which motivates our classification strategy, 
namely: (\textit{i}) to determine the topological space of the Hamiltonians pointwisely in $k$ space, and  (\textit{ii}) to classify the topological configurations of $\mathcal{H}(k)$ on the  
${d_c}$-dimensional sphere  $S^{d_c}$ enclosing the gapless region. 
Here,  $d_c=d-d_{FS}-1$ for a $d_{FS}$-dimensional Fermi surface in $d$-dimensional $k$ space. 
These two steps can be performed by use of the theory of \textit{real} Clifford algebras and  $KO$ theory, respectively. The details of these mathematical derivations are given in the Supplemental Material (SM) \cite{Supp}. Because of the anti-unitarity of $\A$, we treat the imaginary unit $i$ as an operator \cite{Furusaki-K, Supp}, 
which allows us to construct a Clifford algebra with the three generators 
$\A$, $i \A$, and  $i \mathcal{H}$ and the anti-commutators
\begin{equation}
\{\hat{A},i\hat{A}\}=0,\quad \{\hat{A},i\mathcal{H}\}=0,\quad \{i\mathcal{H},i\hat{A}\}=0. \label{Cl-T}
\end{equation}
Assuming that the chemical potential $\mu=0$, it is sufficient for our topological purpose to study flattened Hamiltonians $\tilde{\mathcal{H}}(k)$,
whose band spectra are normalized such that $\tilde{\mathcal{H}}^2(k)=1$  for every point $k$ where the spectrum is gapped.
With this normalization, the squares of the above generators are given by
\begin{equation}
(i\mathcal{H})^2=-1,\quad \hat{A}^2=(i\hat{A})^2=\pm 1, \label{Norm-T}
\end{equation}
where $\hat{A}^2=(i\hat{A})^2$ due to the anti-unitarity of $\hat{A}$. 

We first consider $\A^2=+1$, 
in which case both $\A$ and $i\A$ are positive, generating the Clifford algebra $C^{0,2}$, which is further extended to  $C^{1,2}$ by the negative generator $i\mathcal{H}$. 
In other words, the topological space of all $\PP\TT$ invariant  $\mathcal{H}$ is determined by all possible Clifford algebra extensions from 
$C^{0,2}$ to $C^{1,2}$.
Since the extension $C^{0,2}\subset C^{1,2}$ is equivalent to $C^{0,0}\subset C^{0,1}$, the topological space is  equivalent to $R_0$, the zeroth classifying space of   $KO$ theory~\cite{Atiyah-KR,Karoubi-book,Clifford-modules}. According to our classification strategy, we next need to classify the topological configurations of $R_0$ on the sphere $S^{d_c}$. 
To do so, we note that since the trivial action of $\PP\TT$ in $k$ space corresponds to the trivial involution of $KR$ theory,
the classification is given by $KO$ theory, the simplest instance of   $KR$ theory. Hence, the classification of
$\PP\TT$ invariant Fermi surfaces with $\A^2=+1$ follows from the $KO$ groups
\begin{equation}
\widetilde{KO}(S^{d_c})\cong \mathbb{Z}, \mathbb{Z}_2,\mathbb{Z}_2,0,2\mathbb{Z},0,0,0 \quad ,
\end{equation}
with $\quad d_c\equiv 0,1,\cdots,7 \mod 8$, and $\widetilde{KO}=\widetilde{KO}^{-0}$, where ``$0$'' is determined by the classifying space $R_0$. 

Second, we consider $\hat{A}^2=-1$, in which case   $\hat{A}$ and $i\hat{A}$  generate the Clifford algebra $C^{2,0}$, which is extended by $i\mathcal{H}$ to $C^{3,0}$. 
From $C^{2,0}\subset C^{3,0}\approx C^{0,4}\subset C^{0,5}$, it follows that the topological space of Hamiltonians is given by $R_4$. Thus, the classification of Fermi surfaces with codimension $(d_c+1)$
and $\hat{A}^2=-1$ is  
\begin{equation}
\widetilde{KO}^{-4}(S^{d_c})\cong  2\mathbb{Z},0,0,0,\mathbb{Z}, \mathbb{Z}_2,\mathbb{Z}_2,0\quad ,
\end{equation}
with $ d_c\equiv 0,1,\cdots,7 \mod 8.$

\textit{Topological Fermi surfaces with $\CC\PP$ symmetry.--}
Next, we look into  $\CC\PP$ symmetric Fermi surfaces (or superconducting nodes).
$\CC$ is implemented by an anti-unitary operator $\hat{C}$, which acts on the Hamiltonian as 
\begin{equation}
\hat{C}\mathcal{H}(k)\hat{C}^{-1}=-\mathcal{H}(-k) . 
\end{equation} 
Accordingly, we need to consider the Clifford algebra generated by $\hat{B}=\hat{C}\hat{P}$, $\mathcal{H}$,
and  $i \hat{B}$ with the anti-commutators
\begin{equation}
\{\hat{B},\mathcal{H}\}=0, \quad \{\hat{B},i\hat{B}\}=0,\quad \{\mathcal{H},i\hat{B}\}=0 .
\end{equation}
As before, we normalize the band spectrum
such that
\begin{equation}
\mathcal{H}^2=1,\quad \hat{B}^2=(i\hat{B})^2=\pm 1.
\end{equation}
For $\hat{B}^2=+1$, one can find that the Hamiltonian space is equivalent to $R_{2}$~\cite{Supp}, corresponding to the Clifford algebra extension $C^{0,2}\subset C^{0,3}$. 
From this it follows that the classification is given by
\begin{equation}
\widetilde{KO}^{-2}(S^{d_c})\cong \mathbb{Z}_2,0,2\mathbb{Z},0,0,0,\mathbb{Z}_2,\mathbb{Z}  \quad ,
\end{equation}
with $d_c\equiv 0,1,\cdots,7 \mod 8$. For $\hat{B}^2=-1$, on the other hand, the Hamiltonian space is $R_{6}$, corresponding to $C^{2,0}\subset C^{2,1}\approx C^{0,6}\subset C^{0,7}$. This leads to the classification
\begin{equation}
\widetilde{KO}^{-6}(S^{d_c})\cong 0,0,\mathbb{Z},\mathbb{Z}_2,\mathbb{Z}_2,0,2\mathbb{Z},0 \quad ,
\end{equation}
with $d_c\equiv 0,1,\cdots,7 \mod 8$.
This concludes the derivation of the classification, as summarized in Table~\ref{Classification}.

\begin{table}
	\begin{tabular}{c|cccccccc}
		$d_c$ & $0$ & $1$ & $2$ & $3$ & $4$ & $5$ & $6$ & $7$\\
		\hline  
		$(\P\T)^2=+1$ & $\mathbb{Z}$ & $\mathbb{Z}_2$ & $\mathbb{Z}_2$ & $0$ & $2\mathbb{Z}$ & $0$ & $0$ & $0$\\
		$(\C\P)^2=+1$  & $\mathbb{Z}_2$ & 0 & $2\mathbb{Z}$ & 0 & 0 & 0 & $\mathbb{Z}$ & $\mathbb{Z}_2$\\
		$(\P\T)^2=-1$ & $2\mathbb{Z}$ & $0$ & $0$ & $0$ & $\mathbb{Z}$ & $\mathbb{Z}_2$ & $\mathbb{Z}_2$ & $0$\\
		$(\C\P)^2=-1$ & $0$ & $0$ & $\mathbb{Z}$ & $\mathbb{Z}_2$ & $\mathbb{Z}_2$ & $0$ & $2\mathbb{Z}$ & $0$
	\end{tabular}
	\caption{Classification table of $\PP\TT$ and $\CC\PP$ invariant Fermi surfaces. \label{Classification}}
\end{table}

Several comments are in order. First, note that all classifications are given by $\widetilde{KO}^{-q}(S^{d_c})$, 
where $q$ is even and denotes the index of the classifying space $R_q$ \cite{Eight-classes}.
Second, in the present classification, the $K$ groups as a function of $d_c$ appear in reverse order 
compared to the tenfold classification of Fermi surfaces and strong TIs/TSCs~\cite{KR-KO}. 
Third, if we replace $d_c$ by the spatial dimension  $d$,   $KO$ theory yields the classification of $\PP\TT$ and $\CC\PP$-invariant TIs/TSCs, since $\PP\TT$ and $\CC\PP$ act trivially in  momentum space~\cite{Ryu2010ten}. Note, however, that these TIs/TSCs do not exhibit any
symmetry-protected surface states, since the boundary breaks $\PP\TT$($\CC\PP$).
Finally, we emphasize that the classification relies only on the combined symmetry $\PP\TT$($\CC\PP$). Both $\PP$ and $\TT$($\CC$) may be broken individually, but the combination
must be preserved, which is in contrast to   previous studies of $\PP$-symmetric systems~\cite{Sato_Crystalline_PRB14}.


\textit{Representative models.--}
We now construct representative models for all nontrivial cases of the classification in the physical dimensions $d=1,2$, and $3$.
(The computation of the corresponding topological charges is presented in the SM~\cite{Supp}.)
Discussing the symmetry classes in the same sequence as before, we start
with the symmetry class $\A^2= (\PP\TT)^2 =+1$, for which there exist topologically nontrivial 
Fermi surfaces in all physical dimensions, since for $d_c=0,1$, and $2$ the classification is given 
by $\mathbb{Z}$, $\mathbb{Z}_2$, and $\mathbb{Z}_2$, respectively. 
The case $d_c=0$ simply corresponds to Fermi surfaces of spinless free fermions in any dimension $d$, see Refs.~\cite{Z-charge,Supp}. 
The case $d_c=1$ corresponds to topological Fermi points (lines) in 2D  (3D) with $\mathbb{Z}_2$ topological charge.
For 2D, a simple model can be constructed with rank-$2$ matrices.
We choose $\T=\K$ and $\P=\sigma_3$, which yields
$\A=\sigma_3\K$ and $[\T,\P]=0$, where $\K$ denotes the complex conjugate operator and $\sigma_j$'s are Pauli matrices. (Alternatively one can consider $\T=-i\sigma_2\K$ and $\P=\sigma_1$ with the anti-commutation relation $\{\T,\P\}=0$.)
Thus the general form of the Hamiltonian is 
$\mathcal{H}_0=f_1 (k)\sigma_2+f_2(k)\sigma_3$, with $f_i$ arbitrary functions of $k$, since $\PP\TT$ merely forbids the $\sigma_1$ term.
For concreteness, let us choose
\begin{equation}
\mathcal{H}_0=k_x\sigma_2+(k_y^2-R^2)\sigma_3,  \label{2D-PT}
\end{equation}
with the constant $R\sim 1$. $\mathcal{H}_0$ exhibits two Dirac points at $k= (0, \pm R)$  
with topological charges $\nu=\pm 1$.
The low-energy physics in the vicinity of these two gapless points is described by 
the Dirac-type Hamiltonian $\mathcal{H}_{\pm}=k_x\sigma_2\pm k_y\sigma_3$, whose stability is guaranteed by a quantized Berry phase \cite{Supp,Note-ZhangJ}. 
Note that $\mathcal{H}_0$ has both $\TT$ and $\PP$ symmetries. However,
the Dirac points cannot be gapped out by 
 $\PP$ and $\TT$ breaking perturbations that satisfy  $\PP\TT$, such as
 $\mathcal{H}'=\mu\sigma_0+(\eta+\epsilon_1 k_x^2)\sigma_2+\epsilon_2 k_y\sigma_3 . $ 
These perturbations only change
 the local properties of the Dirac points, e.g., position and dispersion, 
 but do not open a full gap. 
We emphasize that the $\mathbb{Z}_2$ Fermi points of $\mathcal{H}_0$ are fundamentally different from those with $\mathbb{Z}$ classification in symmetry class AIII,
although in both cases $\sigma_1$ terms are symmetry forbidden.
 To illustrate the $\mathbb{Z}_2$ nature of the Fermi points, we consider a doubled version of $\mathcal{H}_0$, namely $\mathcal{H}_0 \otimes \tau_0$.
 It is found that there are $\PP\TT$-preserving perturbations, for instance $m\sigma_1\otimes \tau_2$, that open up a full gap.
 However, all of these gap opening perturbations are forbidden by chiral symmetry. 
 Hence, the discussed  $\mathbb{Z}_2$  Fermi points are clearly distinct 
 from the $\mathbb{Z}$ Fermi points of  class AIII~\cite{matsuuraNJP13,ZhaoWangPRL13,ZhaoWangPRB14}. A lattice version of Eq.~(\ref{2D-PT}), $\mathcal{H}(k)=\sin k_x\sigma_2+(\lambda-\cos k_y)$ ($|\lambda|<1$), will be discussed in detail later. It is also noted that $\mathcal{H}_0$ can straightforwardly be extended to a $3$D case with a nontrivial nodal loop~\cite{Supp}.

Next, we consider the case $\A^2 =+1$ with $d_c=2$. According to Table~\ref{Classification}, there exist in this
symmetry class $\PP\TT$ preserving Fermi points in $d=3$ with a $\mathbb{Z}_2$ charge.
A minimal model for these $\mathbb{Z}_2$ Dirac points can be constructed by rank-$4$ matrices. 
Choosing $\A=\sigma_3 \otimes \tau_0 \K$, we find that the following continuum model exhibits such a Dirac point
\begin{equation} \label{Dirac_ham}
\mathcal{H}_{D}(k)= k_x\sigma_1\otimes\tau_2+k_y\sigma_2\otimes\tau_0+k_z\sigma_3\otimes\tau_0,
\end{equation}
where $\tau_j$'s are a second set of Pauli matrices. Observe that the two independent  mass matrices $\sigma_1\otimes\tau_3$ and $\sigma_1\otimes\tau_1$ are forbidden by   $\PP\TT$ symmetry. However, 
the Fermi point of the doubled Hamiltonian $\mathcal{H}_D\otimes\kappa_0$ can be gapped out by  the mass terms $m\sigma_1\otimes\tau_1\otimes\kappa_2$ and $m\sigma_1\otimes\tau_3\otimes\kappa_2$ with $\kappa_j$'s being Pauli matrices, which illustrates the $\mathbb{Z}_2$ nature of the Dirac point of $\mathcal{H}_D$.

For  symmetry class $\A^2=-1$, the only nontrivial case in $d=1$, $2$, or $3$ is $d_c=0$, which has a $2\mathbb{Z}$ classification (Table~\ref{Classification}).
This simply corresponds to spinful free fermions.
Choosing $\T=i\sigma_2\K$ and $\P=\sigma_0$, which yields $\A=i\sigma_2\K$ and $[\T,\P]=0$, the spinful free-fermion Hamiltonian is given by
$\mathcal{H}_{free}^{s}=k^2/2m\sigma_0-\mu\sigma_0$. Note that all spin-orbit coupling terms involving $\sigma_j$ ($j=1,2,3$) are excluded by  $\PP\TT$ symmetry.

Let us now turn to nodal band structures with $\CC\PP$ symmetry.
For symmetry class $\B^2=(\CC\PP)^2  =+1$ with $d_c=0$ there exist nodes with a $\mathbb{Z}_2$ classification.
To construct a representative model, we 
choose $\C=\tau_1\K$ and $\P=i\tau_2$, so that $\B=\tau_3\K$.  
Since the only Pauli matrix that anti-commutes with $\B$ is $\tau_1$, we find that
 the continuum model is  $\mathcal{H}(k)=k\tau_1$ in 1D.
  It is obvious that the mass terms  $m\tau_2$ and $m\tau_3$ are forbidden by  $\CC\PP$ symmetry. To see the $\mathbb{Z}_2$ nature of the Fermi point, we 
  observe that the doubled Hamiltonian $\tilde{\mathcal{H}}(k)=k\tau_1\otimes\sigma_0$ with the trivial $\mathbb{Z}_2$ charge can be gapped by the $\CC\PP$ invariant terms $m\tau_3\otimes \sigma_2$ and $m\tau_2\otimes\sigma_2$. Models
of these $\CC\PP$ invariant  nodes  in $2$D and $3$D can be constructed 
 in an analogous manner. 

For $\B^2=+1$ and $d_c=2$, there is a $2\mathbb{Z}$ classification. A representative model can be  constructed by $4\times 4$ matrices. 
Choosing $\B=\tau_3 \otimes \sigma_0\K$, the continuum Hamiltonian is given by
\begin{equation}
\mathcal{H}_W^{double}=k_x\tau_1\otimes\sigma_1+k_y\tau_0\otimes\sigma_2+k_z\tau_1\otimes\sigma_3, \label{double-Weyl}
\end{equation}
since there are  only  three mutually anti-commuting $4\times 4$ matrices that also anti-commute with $\B$.
$\mathcal{H}_W^{double}$ exhibits a double Weyl point at $k=0$ with
 topological charge $\nu=2$, which is defined in terms of a Chern number on a sphere enclosing the Weyl point~\cite{Supp,Note-Double-Weyl}.
 Despite its similar appearance, this model should be distinguished from the Dirac Hamiltonian~(\ref{Dirac_ham})  that has a vanishing Chern number.
The remaining nontrivial case in physical dimensions is $d_c=2$ for $\B^2=-1$, which may be exemplified by a Weyl point \cite{Supp}. 

In passing, we note that the $A$ phase of $^3$He can   be viewed as an example of $\CC \PP$ symmetry protected nodal points.
$^3$He-$A$ has both $\CC$ and $\PP$ symmetry with $\C=i\sigma_2\otimes i\tau_2\K$ and $\P=\tau_3$, respectively, which yields $(\C\P)^2=-1$ \cite{Volovik:book}.
Thus, according to Table~\ref{Classification} the classification of the $^3$He-$A$ point nodes is of $\mathbb{Z}$ type with a topological charge
of $\pm 2$, due to spin degeneracy. $\CC\PP$ invariant perturbations, such as spin-orbit coupling,
may split the spin degeneracy, which divides the doubly charged Weyl points into  Weyl points with charge one. This is in contrast to the elementary Fermi point of Eq.~(\ref{double-Weyl}), which has a $2\mathbb{Z}$ classification.

\begin{figure}
	\centering
	\includegraphics[scale=0.85]{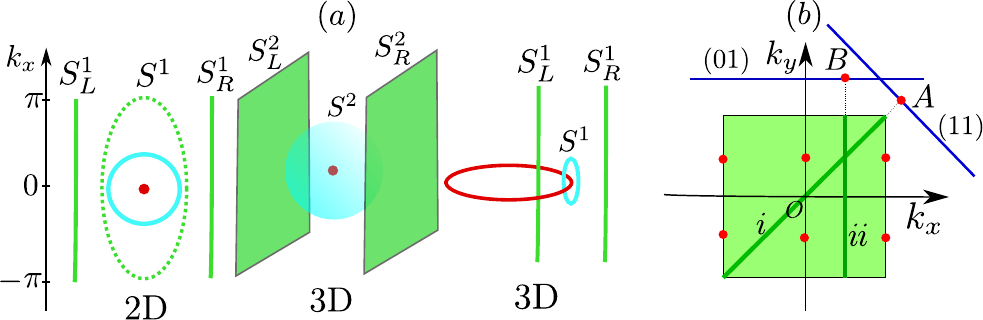}
	\caption{(a) A small sphere/circle (cyan) enclosing the gapless point/ring (red) can be deformed into two large spheres/circles (green) in the  BZ. Note that  the BZ is periodic in $k$. 
	(b) Lattice model with topological Fermi points (red) in its  bulk BZ (light green). The boundary BZs  for the (11) and (01) edges are indicated by blue lines.\label{M-index} 
	}
\end{figure}

\textit{Bulk-boundary correspondences.--}
The distribution of subgap states at the boundary of topological metals/semimetals is determined by the topological charges of the Fermi surfaces in the bulk.
This is illustrated in Fig.~\ref{M-index}(a), which shows how a small sphere $S^2$ (or circle $S^1$) enclosing the bulk gapless region may be deformed continuously into  two large $S^2$'s ($S^1$'s) in the Brillouin zone (BZ) due to the periodicity in $k$.  
By regarding these large $S^2$'s ($S^1$'s)  as subsystems of the whole system, we obtain the following relation
\begin{equation} \label{index_nu_R_L}
\nu=N_{R}-N_{L},
\end{equation}
where $\nu$ is the topological charge of the Fermi surface, and $N_{R/L}$ are the topological numbers on the right/left $S^2$ ($S^1$). 
Since  $\PP\TT$ ($\CC\PP$) acts trivially in $k$, 
all three topological indexes $N_{R}$, $N_{L}$, and $\nu$ belong to the same symmetry class of Table~\ref{Classification}. 
Therefore, Eq.~(\ref{index_nu_R_L}) determines the number of subgap modes on the boundaries that are perpendicular to the subsystems $S^2_{R/L}$ ($S^1_{R/L}$).
In general, there may exist several topologically charged gapless regions in the BZ, which leads to a set of equations of the form~(\ref{index_nu_R_L}) that determines
the distribution of the boundary modes \cite{Note-BBC}.

To illustrate the above bulk-boundary correspondence, we discuss the gapless modes on the (11) and (01) edges of the lattice model of Eq.~(\ref{2D-PT}) [see Fig.~\ref{M-index}(b)]. Consider point $A$ on the (11) edge BZ, which can be viewed as an end point of the $1$D   subsystem $i$. Since $i$ encloses an odd number of nontrivial $\mathbb{Z}_2$ Fermi points, it has a nontrivial topological index given by the geometric phase of the Berry connection. Hence, a subgap state appears at the end point $A$.  In contrast, no edge state
appears at  point $B$ of the (01) edge BZ, since $B$ is the end point of the subsystem $ii$, which encloses an even number of nontrivial $\mathbb{Z}_2$ Fermi points leading to a trivial topological index. This bulk-boundary correspondence can be
utilized in experiments   
to identify  $\PP\TT$ ($\CC\PP$) invariant    materials.  In the SM~\cite{Supp} we present an explicit calculation of these topological indices, confirming the above analyses \cite{SQS-Book}.
 
In closing, we note that by use of a bulk-boundary correspondence analysis, the topological Fermi surfaces of Table~\ref{Classification} can be interpreted as gapless boundary modes of fully gapped TIs/TSCs with certain crystalline symmetries. To demonstrate this, let us consider
a gapless $d$D boundary of a $(d+1)$D fully gapped TI/TSC. 
Since the gapped bulk of the  TIs/TSCs  provides a physical ultraviolet cutoff for the gapless boundary modes, the nontrivial topological configuration  of the TIs/TSCs imprints itself on the ultraviolet behavior of the boundary modes.
To make this more explicit, let us assume that the gapless boundary modes are Fermi points with codimension $d_c =d -1$. The topological charge of these
Fermi points is determined by invariants that are defined on $(d-1)$D spheres $S^{d-1}$  enclosing the Fermi points.
In order to establish a correspondence between the topological charge of the boundary modes and the bulk topology of the $(d+1)$D TIs/TSCs, one must show that
the two have the same classifications, i.e., that the two classifications are mapped onto each other by a two-dimension shift. 
For the strong TIs/TSCs of the tenfold way, such a two-dimension shift arises since the involution due to  $\TT$ or $\CC$ is different for 
the $S^{d_c}$  spheres enclosing the boundary modes and the bulk $k$ space of  the TIs/TSCs \cite{ZhaoWangPRB14,zhao_PRB14b,Supp}.
For $\PP\TT$ and $\CC\PP$ symmetries, however, there is no such involution difference between $S^{d_c}$ and the bulk $k$ space of the TIs/TSCs. 
Hence,  the $\PP\TT$ ($\CC\PP$) invariant Fermi surfaces of Table~\ref{Classification} must be related to TIs/TSCs with a symmetry different from 
$\PP\TT$ ($\CC\PP$). Indeed, we find that the classification of Table~\ref{Classification} is related to the classification of
 $(d+1)$D gapped band structures with the anti-unitary symmetries  $\DD=\TT\RR$ and $\EE=\CC\RR$, where $\RR$ acts
 on the TIs/TSCs as
\begin{equation}\label{new_refl}
\hat{R}\mathcal{H}_{TI}(k,k_{d+1}) \hat{R}^{-1}=\mathcal{H}_{TI}(-k,k_{d+1}),
\end{equation}
and $\DD$ and $\EE$ restrict $\mathcal{H}_{TI}(k,k_{d+1})$ as
\begin{eqnarray} \label{new_syms}
\D\mathcal{H}_{TI}(k,k_{d+1}) \D^{-1}&=&\mathcal{H}_{TI}(k,-k_{d+1}), \nonumber\\ 
\E\mathcal{H}_{TI}(k,k_{d+1}) \E^{-1}&=&-\mathcal{H}_{TI}(k,-k_{d+1}).
\end{eqnarray}
Note that $\RR$ projected onto the boundary acts like an inversion symmetry in the boundary BZ.
The classification of TIs/TSCs with symmetry~(\ref{new_syms}) is given by $KR$ theory as
$KR^{-q}(B^{1,d},S^{1,d})$. It follows from the relation~\cite{Atiyah-KR,Atiyahbook}
\begin{equation}
KR^{-q}(B^{1,d},S^{1,d})\cong \widetilde{KO}^{-q}(S^{d-1})
\end{equation}
that $\PP\TT$ ($\CC\PP$) symmetric Fermi surfaces with codimension $d_c = d-1$ have the same classification as
$(d+1)$D TIs/TSCs with symmetry~(\ref{new_syms}), which establishes the promised exact bulk-boundary correspondence.

\begin{acknowledgments}
We thank A. J. Leggett for helpful discussions. This work  was partially supported by the GRF of Hong Kong (Grants No. HKU173051/14P and No. HKU173055/15P).
\end{acknowledgments}

  \clearpage
  \newpage

  \appendix
  
  
  
  
  \begin{center}
  	\textbf{
  		\large{Supplemental Material for}}
  	\vspace{0.2cm} 
  	
  	\textbf{
  		\large{
  			``Unified theory of $\PP\TT$ and $\CC\PP$ invariant topological metals and nodal superconductors" } 
  	}
  \end{center}
  
  \vspace{-0.2cm}

  \section{Real Clifford algebras}
  We here introduce  a basic knowledge of real Clifford algebras in a causal way. The main contents in this section and the next section come from Refs.\cite{Clifford-modules,Karoubi-book}, which are regarded in a self-contained way for our purpose.
  
  We assume that $e_j$, $j=1,\cdots,p+q$ satisfy the anti-commutation relations,
  \begin{equation}
  	\{e_j,e_k\}=2\eta_{jk},
  \end{equation}
  where $\eta_{jk}$ is defined by the quadratic form
  \begin{equation}
  	Q(x)=-\sum_{j=1}^{p}x_j^2+\sum_{k=1}^q x_{p+k}^2=\eta^{jk} x_j x_k.
  \end{equation}
  Although $e_j$ are mathematically abstract symbols, for our purpose we can simply regard them as a set of real matrices with the commutators above being the matrix commutators. 
  
  Then the real Clifford algebra is a $2^{p+q}$-dimensional real vector space with the basis
  \begin{equation}
  	e_{i_1}e_{i_2}\cdots e_{i_k}, \quad 0<i_1<i_2<\cdots i_k\le p+q,
  \end{equation}
  and the multiplication rule of two vectors are specified by the anti-commutation relations.
  
  For example, the real numbers $\mathbb{R}$ is isomorphic to $C^{0,0}$, $\mathbb{R}\cong C^{0,0},$ and the complex number $\mathbb{C}\cong C^{1,0}$, where the imaginary unit $i$ is the only generator with $i^2=-1$. 
  
  Let $M_{n}(\mathbb{R})$ be the algebra of $n\times n$ real matrices. Then \begin{equation}
  	M_{2}(\mathbb{R})\cong C^{1,1}\cong C^{0,2}.
  \end{equation}
  To see this, we introduce
  \begin{equation}
  	I=\begin{pmatrix}
  		0 & -1\\
  		1 & 0
  	\end{pmatrix}, K=\begin{pmatrix}
  	1 & 0\\0 & -1
  \end{pmatrix}, J=\begin{pmatrix}
  0 & 1\\ 1 & 0
\end{pmatrix},
\end{equation}
which satisfy the relations
\begin{equation}
	IK=J,~~JK=I,~~-I^2=J^2=K^2=-1_2
\end{equation}
with $1_2$ being the identity matrix. If we choose $I$ and $K$ as generators, then $M_2(\mathbb{R})\cong C^{1,1}$, and if we choose $K$ and $J$, $M_2(\mathbb{R})\cong C^{0,2}$.

We give our last example without proof, which is
\begin{equation}
	\begin{split}
		M_{16}(\mathbb{R})&\cong C^{0,8}\cong C^{0,4}\otimes C^{0,4}\\
		&\cong C^{4,0}\otimes C^{0,4}\cong C^{4,0}\otimes C^{4,0}\cong  C^{8,0}.
	\end{split}
\end{equation}

\section{Eight classifying spaces}
Given a real Clifford algebra $C^{p,q}$, we may add one more positive generator to extend $C^{p,q}$ to $C^{p,q+1}$, which is denoted as $C^{p,q}\subset C^{p,q+1}$, or a negative one corresponding to $C^{p,q}\subset C^{p+1,q}$. All possible extensions of $C^{p,q}\subset C^{p,q+1}$ or $C^{p,q}\subset C^{p+1,q}$ for given $C^{p,q}$ form a topological space. As we will see, there are only eight independent such topological spaces for all one-generator extensions of Clifford algebras, and all of which can be enumerated by $C^{0,q}\subset C^{0,q+1}$ with $q=0,1,\cdots,7$. The eight spaces are called classifying spaces of the $KO$ theory.

To see this, we need the following equivalent relations of Clifford algebras,
\begin{eqnarray}
	C^{p,q}\otimes M_2(\mathbb{R}) &\cong& C^{p+1,q+1}\\ \label{one-one}
	C^{p,q}\otimes M_2(\mathbb{R}) &\cong& C^{q,p+2}\\ \label{exchange}
	C^{p,q}\otimes M_{16}(\mathbb{R}) &\cong& C^{p,q+8} \label{eight-fold}
\end{eqnarray}
which, respectively, lead to the equivalence relations of the extensions of Clifford algebras,
\begin{eqnarray}
	C^{p,q}\subset C^{p,q+1} &\cong& C^{p+1,q+1} \subset C^{p+1,q+2}\\
	C^{p,q}\subset C^{p+1,q} &\cong& C^{q,p+2} \subset C^{q,p+3}\\
	C^{p,q+8}\subset C^{p,q+9} &\cong& C^{p,q} \subset C^{p,q+1}.
\end{eqnarray}
The above three equivalence relations are sufficient to prove our conclusion about the eight independent extension spaces. Actually they are also used for several times in our main text to work out the space of the Hamiltonians under the restriction of symmetries. Our remaining task is to prove Eqs.(\ref{one-one}),(\ref{exchange}) and (\ref{exchange}).

We now prove Eq.(\ref{eight-fold}). Let $\beta_1,\cdots,\beta_8$ be generators of $C^{0,8}$, and introduce $\epsilon=\beta_1\beta_2\cdots\beta_8$, which satisfy $\epsilon^2=1$ by straightforward calculation. Recalling that $C^{0,8}\cong M_{16}(\mathbb{R})$, we may choose elements of $C^{p,q}\otimes M_{16}(\mathbb{R})$, 
\begin{equation}
	e_1\otimes \epsilon, \cdots, e_{p+q}\otimes \epsilon, 1\otimes\beta_1,\cdots,1\otimes\beta_8,
\end{equation}
which form a complete set of generators of $C^{p,q+8}$. The other two can be proved in a similar way. In particular, we choose $\epsilon=IK$ for Eq.(\ref{one-one}), and $\epsilon=KJ$ for Eq.(\ref{exchange}).

We denote the eight classifying spaces by $R_q$, $q=0,\cdots,7$, where $R_q$ corresponds to the extension $C^{0,q}\subset C^{0,q+1}$. Explicitly these spaces are presented in Tab.\ref{Classifying-spaces}.

\begin{table*}
	\begin{tabular}{c|cccccccc}
		$q$ & 0 & 1 & 2 & 3 & 4 & 5 & 6 & 7\\
		\hline  
		$R_{q}$ & $\frac{O(M+N)\times\mathbb{Z}}{O(M)\times O(N)}$ & $O(N)$ & $\frac{O(2N)}{U(N)}$ & $\frac{U(2N)}{Sp(N)}$ & $\frac{Sp(M+N)\times\mathbb{Z}}{Sp(M)\times Sp(N)}$ & $Sp(N)$ & $\frac{Sp(N)}{U(N)}$ & $\frac{U(N)}{O(N)}$\\
		$\pi_0(R_q)$ & $\mathbb{Z}$ & $\mathbb{Z}_2$ & $\mathbb{Z}_2$ & 0 & $2\mathbb{Z}$ & 0 & 0 & 0
	\end{tabular}
	\caption{The eight real classifying spaces and their zeroth homotopy groups. \label{Classifying-spaces}}
\end{table*}

\section{Mapping complex matrices to real ones}
In the main text,  we treat essentially Hamiltonians as real matrices under additional restrictions. Now we justify this point. It is observed that $\mathbb{C}\cong C^{1,0}\subset C^{1,1} \cong M_2(\mathbb{R})$, which suggests us to embed $\mathbb{C}$ into $M_2(\mathbb{R})$. Explicitly, since $I^2=-1_2$, we map 
\begin{equation}
	1\longmapsto 1_2,\quad i\longmapsto I,
\end{equation}
which leads to $z=x+iy\longmapsto x 1_2+ yI$, defining the mapping from $\mathbb{C}$ to $M_2(\mathbb{R})$. This map is obvious injective, and its image set is specified by the condition,
\begin{equation}
	[M,I]=0, \quad M\in M_2(\mathbb{R}),
\end{equation}
which justifies why in the main text we treat the imaginary image $i$ as an operator. The complex conjugate operator $\mathcal{K}$ is now represented in $M_2(\mathbb{R})$ by $K$ (or $J$), operating as
\begin{equation}
	K I K=-I.
\end{equation}
To map a Hamiltonian $\mathcal{H}(k)$ to be a real matrix, we just map all its complex entries to be real matrices with the size of the matrix being doubled. 

\section{The computation of K groups}
In this section, we give calculation details of the $K$ groups in the main text. All of them are done by the $KR$ theory, since the $KO$ theory is just the $KR$ theory with trivial involution in the base space. All the mathematics in this section can be found in Atiyah' original paper on the $KR$ theory \cite{Atiyah-KR}. We follow Atiyah' notation that $\mathbb{R}^{p,q}$ denote a $(p+q)$-dimensional vector space $\mathbb{R}^p\oplus \mathbb{R}^q$ with an involution, $(x,y)\longrightarrow (-x,y)$, with $x\in \mathbb{R}^p$ and $y\in \mathbb{R}^q$. $B^{p,q}$ and $S^{p,q}$ are, respectively, the unit solid ball and the unit $(p+q-1)$-dimensional sphere in $\mathbb{R}^{p,q}$ with the induced involution.

The $KO$ groups for all the classifications of $\TT\PP$ and $\CC\PP$ invariant Fermi surfaces are computed as
\begin{equation}
	\begin{split}
		&\quad~\widetilde{KO}^{-q}(S^{d_c}) \\
		&= KR^{-q}(B^{0,d_c}, S^{0,d_c})\\
		&\cong KR(B^{0,d_c}\times B^{0,q}, S^{0,d_c}\times B^{0,q} \cup S^{0,q}\times B^{0,d_c})\\
		& \cong KR(B^{0,d_c+q}, S^{0,d_c+q})\\
		& \cong KR^{0,d_c+q}(pt)\\
		& \cong \pi_{0}(R_{d_c+q})\quad,
	\end{split}
\end{equation}
where `$pt$' denotes a point, and $\pi_{0}(R_{d_c+q})$ can be read from Tab.\ref{Classifying-spaces}.

The $KR$ groups for the classification of the bulk topological insulators/superconductors in the bulk-boundary correspondence in the main text are computed similarly as

\begin{equation}
	\begin{split}
		&\quad~ KR^{-q}(B^{1,d}, S^{1,d}) \\
		&\cong KR(B^{1,d}\times B^{0,q}, S^{1,d}\times B^{0,q} \cup S^{0,q}\times B^{1,d})\\
		& \cong KR(B^{1,d+q}, S^{1,d+q})\\
		& \cong KR^{1,d+q}(pt)\cong KR^{-(d+q-1)}(pt)\\
		& \cong \pi_{0}(R_{d+q-1})\quad .
	\end{split}
\end{equation}
If $d_c=d-1$, then obviously
\begin{equation}
	KR^{-q}(B^{1,d}, S^{1,d})=\widetilde{KO}^{-q}(S^{d-1}) .
\end{equation}

\section{Other models in the classification}
The $\PP\TT$ invariant $3$D model mentioned in the main text is given by
\begin{equation}
	\mathcal{H}_0(k)=[\kappa^2-\alpha(k_x^2+k_y^2)-\alpha_z k_z^2]\sigma_3+\beta k_z\sigma_2,
\end{equation}
which has both $\TT$ and $\PP$ symmetries, and the corresponding gapless points form a circle in the momentum space. $\TT$ and $\PP$ breaking but $\TT\PP$ preserving perturbations, such as
\begin{equation}
	\mathcal{H}'(k)=[\lambda(k_x+k_y)+\lambda_z k_z]\sigma_3+[\eta+\epsilon_z k_z^2+\epsilon (k_x^2+k_y^2)]\sigma_2,
\end{equation}
can deform the nodal loop and shift its position, but can never gap the system.

For the case of $(\C\P)^2=\B^2=-1$ and $d_c=2$, we may choose $\C=\tau_1\K$ and $\P=\tau_3$, so that $\B=i\tau_2\K$, which anti-commutes with all three $\tau_j$. The coarse-gained model is just a Weyl point,
\begin{equation}
	\mathcal{H}_{W}=k_x\tau_x+k_y\tau_y+k_z\tau_z.
\end{equation}

\section{The computation of topological charges}
In this section we provide calculation details of the topological charges in the main text.

For $d$D ideal fermions, $\mathcal{H}=\frac{p^2}{2m}-\mu$, we may choose a circle in the $(\omega,k)$ space to enclose the $(d-1)$D Fermi surface from its transverse dimensions, then the integer topological charge is given as
\begin{equation}
	\nu_{\mathbb{Z}}=-\frac{1}{2\pi i}\oint_{S^{1}} G(\omega,k) d G^{-1}(\omega,k),
\end{equation} 
where $G=1/(i\omega-\mathcal{H}(k))$ is the imaginary Green's function. If we parametrize the circle by $\phi\in[0,2\pi)$, then
\begin{equation}
	\begin{split}
		\nu_{\mathbb{Z}} &=-\frac{1}{2\pi i}\int_0^{2\pi}d\phi ~(i\cos\phi-\sin\phi)^{-1} \partial_{\phi} (i\cos\phi-\sin\phi)\\
		&=1.
	\end{split}               
\end{equation} 
Restricted in the $k$ space, the circle is just an $S^0$ consisting of two points separated by the Fermi surface, and the integer topological charge above may be regarded as the zeroth Chern number on the $S^{0}$.

For the coarse-gained model $\mathcal{H}(k)=k_x\sigma_x+k_y\sigma_y$, we still choose an $S^{1}$ enclosing the gapless point from the gapped region. The the Hamiltonian restricted on the $S^{1}$ is $h(\phi)=k(\cos\phi\sigma_x+\sin\phi\sigma_y)$, whose negative eigenstates are 
\begin{equation}
	|-,\phi\rangle =\frac{1}{\sqrt{2}}\begin{pmatrix}
		1\\
		- e^{i\phi}
	\end{pmatrix}.
\end{equation}
It is noted that $|-,\phi\rangle$ is globally well defined or monotonic over the whole $S^1$. Then the $\mathbb{Z}_2$ topological charge is just the geometric phase accumulated around the circle under proper normalization,
\begin{equation}
	\nu_{\mathbb{Z}_2}= \frac{1}{i\pi}\int_0^{2\pi} d\phi~ \langle -,\phi| \partial_\phi|-,\phi\rangle\equiv 1 \mod 2.
\end{equation}
Equally good eigenstates may be given by large gauge transformations, $|-,\phi\rangle\longrightarrow e^{in\phi}|-,\phi\rangle$, which increase the topological charge by $2n$ in accord with the $\mathbb{Z}_2$ nature of the topological charge. It is also noted that the renormalized geometric phase is just the first Chern-Simons term of the Berry connection.

For the coarse-gained model $\mathcal{H}^{double}_W=k_x\tau_1\otimes\sigma_1+k_y\tau_0\otimes\sigma_2+k_z\tau_1\otimes\sigma_3$ in the $2\mathbb{Z}$ classification, we choose a cylinder $(-\infty,\infty)\times S^2$ in the $(\omega,k)$ space with $\omega\in (-\infty,\infty)$ and $k\in S^{2}$ to enclose the gapless point. Then the topological charge is given by the Chern number in terms of the imaginary Green's function,
\begin{equation}
	\nu_{2\mathbb{Z}}=\frac{1}{24\pi^2} \int d\omega d\theta d\phi~ \epsilon^{\mu\nu\lambda} \mathrm{tr} G\partial_\mu G^{-1} G\partial_\nu G^{-1} G\partial_\lambda G^{-1},
\end{equation}
where the Green's function is restricted on the cylinder, $G^{-1}(\omega,\theta,\phi)=i\omega-h(\theta,\phi)$, with the $S^2$ being parametrized by $\theta\in[0,\pi]$ and $\phi\in(0,2\pi]$. We introduce $h(\theta,\phi)=g^j\Sigma_j$, where $\Sigma_1=\tau_1\otimes\sigma_1$, $\Sigma_2=\tau_0\otimes\sigma_2$ and $\Sigma_3=\tau_1\otimes\sigma_3$, then
\begin{equation}
	\begin{split}
		\nu_{2\mathbb{Z}}=-\frac{i}{8\pi^2}\int_{-\infty}^{\infty} d\omega \frac{1}{(\omega+r^2)^2}&\int d\theta d\phi~ \epsilon^{jk}g^{n}\partial_j g^m\partial_k g^l\\
		&\mathrm{tr} \Sigma_n\Sigma_m\Sigma_l
	\end{split}
\end{equation}
with $r$ being the radius of the $S^2$, which for our model turns out to be $\nu_{2\mathbb{Z}}=2$ since $\mathrm{tr} \Sigma_n\Sigma_m\Sigma_l=2\mathrm{tr} \sigma_n\sigma_m\sigma_l$.

The topological charge of $\mathcal{H}_W=k_x\tau_x+k_y\tau_y+k_z\tau_z$ may be computed parallel to the one previous one, which is $\nu_{\mathbb{Z}}=1$.

\section{The $\PP\TT$ invariant lattice model of $\mathbb{Z}_2$ Fermi points}
In the momentum space, the lattice version of the Eq.(12) in the main text reads 
\begin{equation}
	\mathcal{H}(k)=\sin k_x \sigma_2+(\lambda-\cos k_y)\sigma_3. \label{PT-2D}
\end{equation}
with $|\lambda|<1$.
The corresponding tight-binding model is given by
\begin{equation}
	\begin{split}
		\hat{H}=&\frac{1}{2}\sum_{\mathbf{i}}(a^\dagger_{\mathbf{i}+\mathbf{e}_x,B} a_{\mathbf{i},A}-a^\dagger_{\mathbf{i}+\mathbf{e}_x,A} a_{\mathbf{i},B})+\mathrm{H.c.}\\
		&-\frac{1}{2}\sum_{\mathbf{i}}(a^\dagger_{\mathbf{i}+\mathbf{e}_y,A} a_{\mathbf{i},A}-a^\dagger_{\mathbf{i}+\mathbf{e}_y,B} a_{\mathbf{i},B})+\mathrm{H.c.}\\
		&+\lambda \sum_{\mathbf{i}}(a^\dagger_{\mathbf{i},A} a_{\mathbf{i},A}-a^\dagger_{\mathbf{i},B} a_{\mathbf{i},B}),
	\end{split}
\end{equation}  
where a pair of Eq.(12)'s in the main text appear in the Brillouin zone.

Now we examine the results obtained in the main text from our general theory by directly solving the lattice model. The low energy effective Hamiltonian of the subsystem $i$ is
\begin{equation}
	h_i(k)=\lambda k \sigma_2-[(1-\lambda)-\frac{1}{4}k^2]\sigma_3.
\end{equation}
Since $(1-\lambda)>0$ with opposite sign compared with the coefficient of $k^2$, there are zero mode solution of the corresponding differential equation
\begin{equation}
	[-\lambda i \partial_x\sigma_2-\frac{1}{4}\partial^2_x\sigma_3-(1-\lambda)\sigma_3]\psi(x)=0
\end{equation}
localized on the end \cite{SQS-Book}. We use $x$ as the real space coordinate of the subsystem $i$, which should not be confused with the $x$ direction of the system. On the other hand the subsystem with trivial topological number has the low energy effective theory,
\begin{equation}
	h_{ii}(k)= k_y\sigma_3\tau_3+m\sigma_2,
\end{equation}
with $\tau$ for the two local energy minima.  Since there exists no $k^2$ term, the differential equation
\begin{equation}
	(-i\partial_y\sigma_3\tau_3+m\sigma_2)\psi(y)=0
\end{equation}
has no solution localized at the end point $B$.


\begin{thebibliography}{57}
\bibitem[{KR-()}]{KR-KO}
	\bibinfo{note}{This is due to the involution difference of the symmetry
		operations in momentum space: $\PP\TT$ and $\PP\CC$ symmetries leave $k$
		fixed, whereas $\TT$ and $\CC$ map $k$ to $-k$. The classifications in the
		real Altland-Zirnbauer classes are given by the $KR$ theory with the involution
		of $k$ to $-k$, while the $KO$ theory is just the simplest case of the $KR$
		theory with trivial involution.}
\bibitem[{Eig()}]{Eight-classes}
	\bibinfo{note}{In principle, symmetry classes with odd $q$ may also be defined,
		which would lead to an eightfold periodicity of the classification with
		respect to both $q$ and $d_c$.}
\bibitem[{Z-c()}]{Z-charge}
	\bibinfo{note}{For the classification $\mathbb{Z}$, the topological charge is
		just given by the Chern number, which actually requires no symmetry.}
\bibitem[{Not({\natexlab{a}})}]{Note-ZhangJ}
	\bibinfo{note}{Nodal points protected by $\PP\TT$ symmetry are also realized in
		two-band Hamiltonians with the symmetry operators $\A=\sigma_{1,0}\K$, whose
		low-energy continuum description reads $H_{2,3}=(k_x-K_x) \sigma_1+(k_y-K_y)
		\sigma_{2,3}$, where $\mathbf{K} = (K_1, K_2)$ denotes the position of the
		nodal point. The Hamiltonian $H_2$ describes the two Dirac points in graphene
		($\hat{P}=\sigma_1,\hat{T}=\K$), while $H_3$ has recently been realized in
		the cold atom experiment of Ref.~\cite{ZhangJ-Exp}, although in that work the
		exotic $\PP\TT$-invariant $\mathbb{Z}_2$ topology was not recognized.}
\bibitem[{Not({\natexlab{b}})}]{Note-Double-Weyl}
	\bibinfo{note}{Note that spin-degenerate Weyl points described by
		$\H_W=\sigma_0\otimes\tau\cdot \mathbf{k}$, with the $\CC\PP$ symmetry given
		by $\C\P=\sigma_2\otimes\tau_2 \K$, belong to the same symmetry class as that
		of Eq.~(14). These spin-degenerate Weyl points may be realized in carbon
		allotropes~\cite{FanZhang2}.}
\bibitem[{Not({\natexlab{c}})}]{Note-BBC}
	\bibinfo{note}{Note that the bulk classification of Table~\ref{Classification},
		for which we have assumed that any boundary is at the asymptotic infinity
		(cf.~Ref.~\cite{Volovik:book}), is not spoiled by the $\PP$ symmetry breaking
		at the boundary.}
\end{thebibliography}
\end{document}